\documentclass[a4paper,12pt]{article}

\usepackage[dvips]{graphicx}
\usepackage[font=small,format=plain,labelfont=bf,up,textfont=normal,justification=justified,singlelinecheck=false]{caption}
\usepackage{amssymb}
\usepackage{amsmath}
\usepackage{longtable}
\usepackage{cite}
\usepackage{hyperref}

\usepackage[usenames,dvipsnames,svgnames,table]{xcolor}

\title{Energy Losses of Magnetic Monopoles in Aluminum, Iron and Copper}
\author{S. Cecchini$^{1}$, L. Patrizii$^{1}$, Z. Sahnoun$^{1,2}$, G. Sirri$^{1}$ and V. Togo$^{1}$}
\date{\small {$^{1}$ INFN Sez. Bologna, viale B. Pichat 6/2, Bologna, I-40127, Italy.\\
              $^{2}$ Astrophys. Dept., CRAAG, Algiers, Algeria.}}

\begin{document}
\maketitle

\begin{abstract}
The Energy Losses and Ranges of magnetic monopoles with magnetic charges $1g_{D}$, $2g_{D}$, $3g_{D}$, $6g_{D}$ and $9g_{D}$ in Aluminum, Iron and in Copper are computed, in the different regimes of velocities. The Restricted Energy Losses of monopoles with magnetic charges $1g_{D}$, $2g_{D}$ and $3g_{D}$ in Nuclear Track Detector is also given.
\end{abstract}

\section{Introduction}
\label{sec:intro}

In 1931 Dirac introduced the magnetic monopole (MM) in order to explain the quantization of the electric charge, which follows from the existence of at least one free magnetic charge \cite{dirac}. He established the basic relationship between the  elementary electric charge $e$ and the basic magnetic charge $g$
\begin{equation}
\label{eq:charge}
	eg=n\hbar c/2
\end{equation}
where $n$ is an integer, $n=1,2,..$.\\
\noindent The magnetic charge is $g = n g_{D}$;\ $g_{D}=\hbar c/2e = 68.5 e$ is called the unit Dirac charge.\\

There is no prediction of the Dirac monopole mass. A rough estimate can be made assuming that the classical monopole radius is equal to the classical electron radius: $r_M= \frac{g^{2}}{m_Mc^{2}}= r_e=\frac{e^{2}}{m_ec^{2}}$, from which $m_M=\frac{g^{2}m_e}{e^{2}} \simeq n \ 4700\  m_e \simeq n \ 2.4\  GeV/c^{2}$. Thus the mass should be relatively large and even larger if the basic charge is $e/3$ and if $n>1$. A new type of spherically symmetric monopole was proposed from electroweak theory \cite{cho} with a mass estimated to range between 3 to 7 TeV \cite{chomass} which makes it a very good candidate for searches at the LHC..\\

The main properties of magnetic monopoles are obtained from the Dirac relation~\ref{eq:charge}, and are summarized  here. We recall that the Dirac relation may be easily obtained semiclassically by considering the system of one monopole and one electron, and quantizing the radial component of the total angular momentum~\cite{prof}. \par 
\noindent - {\it Magnetic charge.} 
If $n$~=1 and if the basic electric charge is that of the electron, then  the basic magnetic charge is $ g_D =\hbar c/ 2e=137e/2=3.29\times 10^{-8} \ cgs=\ 68.5 e$. The magnetic charge should be larger if  $n>1$ and also if the basic electric charge is $e/3$.\par
\noindent - {\it Coupling constant.} 
In analogy with the fine structure constant, $\alpha =e^{2}/\hbar c\simeq 1/137$, the  dimensionless magnetic coupling constant is $ \alpha_g=g^{2}_{D}/ \hbar c \simeq 34.25$; notice that it is very large, much larger than 1, and thus perturbative methods cannot be used.
\par
\noindent - {\it Energy W acquired in a magnetic field  B} : $W = ng_{D} B\ell = n \ 20.5$ keV/G~cm, where $\ell$ the coherent length. If $n=1$,  $\ell\simeq 2$ m and $B\simeq 10~$T the energy gained by the monopole is $ W \simeq 0.4$ TeV.
\par                 
\noindent- {\it Trapping of MMs in ferromagnetic materials.}  
MMs may be trapped in ferromagnetic materials by an image force, which could reach the value of $\simeq 10$ eV/\AA.
\par
\noindent- Electrically charged monopoles (dyons) may arise as quantum--mechanical excitations or as M--p, M-nucleus composites.
\par
\vspace{0.2cm} 
\par
The interactions of MMs with matter  are connected with the electromagnetic properties of MMs and thus are consequences of the Dirac relation.\par
\noindent- {\it Energy losses of fast poles.} 
A fast MM with magnetic charge $g_D$ and velocity $v=\beta c$ behaves like an equivalent electric charge $(ze)_{eq}=g_D\beta$; the energy losses of fast monopoles are thus very large.
\par
\noindent - {\it Energy losses of slow monopoles} ($10^{-4}<\beta<10^{-2}$).
 For slow particles it is important to distinguish the energy lost in ionization or  excitation of atoms and molecules of the medium (``electronic'' energy loss) from that lost to yield kinetic energy to recoiling atoms or nuclei (``atomic'' or ``nuclear'' energy loss). Electronic energy loss dominates for electrically or magnetically charged particles with $\beta> 10^{-3}$.  The $dE/dx$ of MMs  with $10^{-4}<\beta<10^{-3}$ is mainly due to excitations of atoms. In an ionization detector using noble gases there would be, for $10^{-4}<\beta<10^{-3}$, an additional energy loss due to atomic energy level mixing (Drell effect).
 \par
\noindent - {\it Energy losses at very low velocities.} 
MMs with   $v<10^{-4}c$ cannot excite atoms; they can only lose energy in elastic collisions with atoms or with nuclei. The energy is released to the medium in the form of elastic vibrations and/or infra--red radiation \cite{derkaoui1}.\par
\begin{figure}[h]
	\begin{center}
		\includegraphics[width=0.8\textwidth]{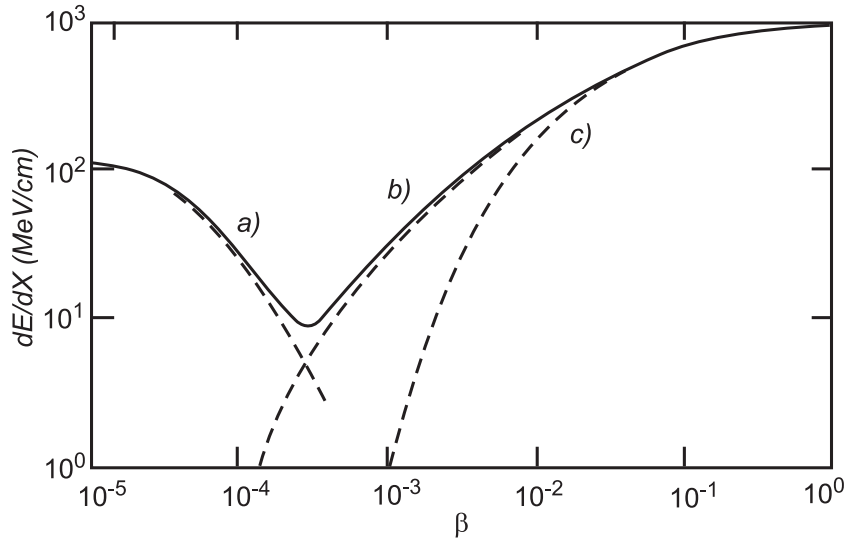}
	\end{center}
	\caption{The energy losses, in MeV/cm, of $g=g_D$ MMs in liquid hydrogen as a function of ${ \beta}$. Curve a) corresponds to elastic monopole--hydrogen atom scattering; curve b) corresponds to interactions with level crossings; curve c) describes the ionization energy loss.}
	\label{fig:perdita-di-energia}
\end{figure}

Fig.\ \ref{fig:perdita-di-energia} shows a sketch of the energy losses in liquid hydrogen of a $g=g_D$ MM vs its $\beta$ \cite{gg+lp}.\par

\noindent - {\it Energy losses in superconductors.}
 If a pole passes through a superconducting ring, there will be a magnetic flux change of $\phi_B=2\pi\hbar c/e$, yielding $dE/dx\simeq 42$~MeV/cm, $\beta-$independent.\par

\section{Energy losses of fast monopoles at $\beta>10^{-1}$ }
\label{sec:fast}

The energy loss of magnetic monopoles with $\beta > 10^{-1}$ was calculated from the Bethe-Bloch formula adapted to magnetic charges \cite{Ahlen} using the following notation
\begin{equation}
\label{eq:fast}
%-\frac{dE}{dx} = \frac{4\pi \, N_{e} \,g^{2} \,e^{2}}{m_{e}c^{2}} \left[\ln\left(\frac{2m_{e}c^{2}\beta^{2}\gamma^{2}}{I} \right) - \frac{1}{2} +\frac{k}{2} - \frac{\delta}{2} - B_{m}\right]
-\frac{dE}{dx} = C \, \frac{Z}{A} \,g^{2} \left[\ln\left(\frac{2m_{e}c^{2}\beta^{2}\gamma^{2}}{I} \right) - \frac{1}{2} +\frac{k}{2} - \frac{\delta}{2} - B_{m}\right] \quad \textrm{MeVg$^{-1}$cm$^2$}
\end{equation}

%$N_{e}$ is the density of electrons, $m_e$ the electron mass, $g$ the magnetic charge of the monopole, $I$ the mean ionization potential, $\delta$ the density effect correction, $k$ the QED correction, and $B_m$ the Bloch correction.\\
\noindent with $C= \frac{e^4}{m_u\,4\pi\epsilon_{0}^{2}  m_e c^2}=0.307$ MeVg$^{-1}$cm$^2$, $m_u$ is the unified atomic mass unit, $m_e$ the electron mass, $g = n g_D = n \cdot 68.5$ the magnetic charge of the monopole, $I$ the mean ionization potential, $\delta$ the density effect correction, $k$ the QED correction, and $B_m$ the Bloch correction.

The QED and Bloch correction terms for different values of the magnetic charge, as given in Derkaoui et al. \cite{derkaoui1}, are

\begin{equation*}
%\label{eq:qed}
 \begin{array}{ll}
k(|g|) = \left\{ \begin{array}{ll} 
0.406 & \textrm{for} \, |g| =137e/2,\\
0.346 & \textrm{for} \, |g| =137e,\\
0.300 & \textrm{for} \, |g| >3 \times 137e/2,\
\end{array} \right.  
%\end{equation}
&
%\begin{equation}
%\label{eq:bloch}
B_m(|g|) = \left\{ \begin{array}{ll}
0.248 & \textrm{for} \, |g| = 137e/2,\\
0.672 & \textrm{for} \, |g| = 137e,\\
1.022 & \textrm{for} \, |g| = 3 \times 137e/2,\\
1.685 & \textrm{for} \, |g| = 6 \times 137e/2,\\
2.085 & \textrm{for} \, |g| = 9 \times 137e/2,\
\end{array} \right.
\end{array} 
\end{equation*}

The density effect term, as given in \cite{pdg}

\begin{equation*}
\delta(\beta\gamma) = \left\{\begin{array}{ll}
2(\ln10)x -\bar{C}                          &   \quad  \textrm{if} \quad x \ge x_1;\\
2(\ln10)x -\bar{C} + a(x_1-x)^k   &    \quad  \textrm{if} \quad x_0 \le x < x_1;\\
0                                                     &    \quad  \textrm{if} \quad x < x_0 (nonconductors);\\
\delta_0 \cdot 10^{2(x-x_0)}       &     \quad \textrm{if} \quad x < x_0 (conductors);\
\end{array} \right.
\end{equation*}

\noindent where $x = \log_{10}(p/Mc)$.\\

Parameters' values used in the computation of the density effect are reported in Table~\ref{tab:delta} \cite{density}:
\begin{table}[h!]
\centering
\begin{tabular}{|c|c|c|c|c|c|c|}
\hline
Material & a & k & $x_0$ & $x_1$ & $\bar{C}$ & $\delta_0$ \\
\hline
Al &   0.08024	& 3.6345 & 0.1708 & 3.0127	& 4.2395 & 0.12 \\
\hline
Fe &   0.14680	& 2.9632 & -0.0012	 & 3.1531 & 4.2911 & 0.12 \\
\hline
Cu &   0.14339	& 2.9044 & -0.0254	 & 3.2792  & 4.4190	& 0.08 \\
\hline
%Si &   0.14921	& 3.2546 &	0.2015 & 2.8716  & 4.4351 & 0.14 \\
%\hline
\end{tabular} 
\caption{Density effect parameters for Aluminum, Iron and Copper as given by Sternheimer~\cite{density}.}
\label{tab:delta}
\end{table}

The mean ionization potential  $I$ of Aluminum, Iron and Copper together with the material's densityl $\rho$ and the electron density  $N_e = \rho\frac{1}{m_u}\frac{Z}{A}$ are given in Table \ref{tab:prop}.\\ 

\begin{table} [h!]
\centering
\begin{tabular}{|c|c|c|c|}
\hline
Material & $\rho$ (g/cm$^3$)   & $N_e$ (cm$^{-3}$)             & $I$(eV) \\
\hline
Al           &  2.699	                  & $7.83 \times 10^{23}$      & $166 \exp(-0.056/2)$ \\
\hline
Fe          &  7.874	                  & $2.2 \times 10^{24}$        & $ 285 \exp(-0.14/2)$ \\
\hline
Cu          &  8.96	                  & $ 2.46 \times 10^{24}$     & $322 \exp(-0.13/2)$ \\
\hline
%Si        &  2.33	                  & $6.99 \times 10^{23}$      & $173$ \\
%\hline
\end{tabular} 
\caption{Density $\rho$, electron density $N_e$ and mean ionization potential $I$ for Aluminum, Iron and Copper.}
\label{tab:prop}
\end{table}

\begin{figure}[h]
          \vspace{0.5cm}
	\begin{center}
		\includegraphics[width=0.8\textwidth]{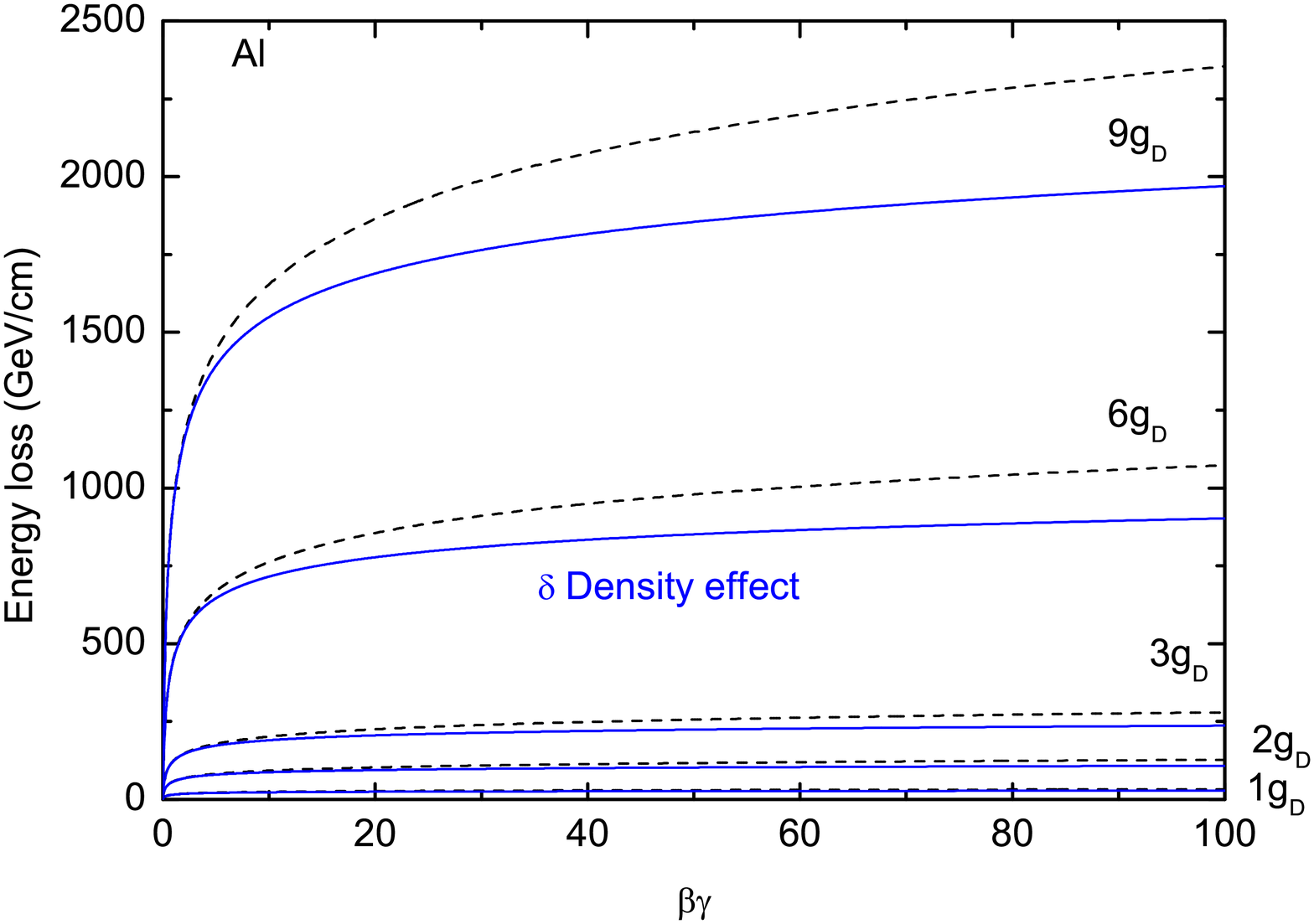}
	\end{center}
	\vspace{-0.5cm}
	\caption{Energy loss of MMs in Aluminum vs $\beta\gamma$ computed with (blue curves) and without (black dashed curves) the density effect correction taken into account. MMs of magnetic charges $1g_D, 2g_D,3g_D, 6g_D$ and $9g_D$ are considered.}
\label{fig:densityAlu1}
\end{figure}

\begin{figure*}[h!]
\vspace{1.0cm}
%	\hspace{-0.2cm}
	\begin{center}
		\includegraphics[width=0.82\textwidth]{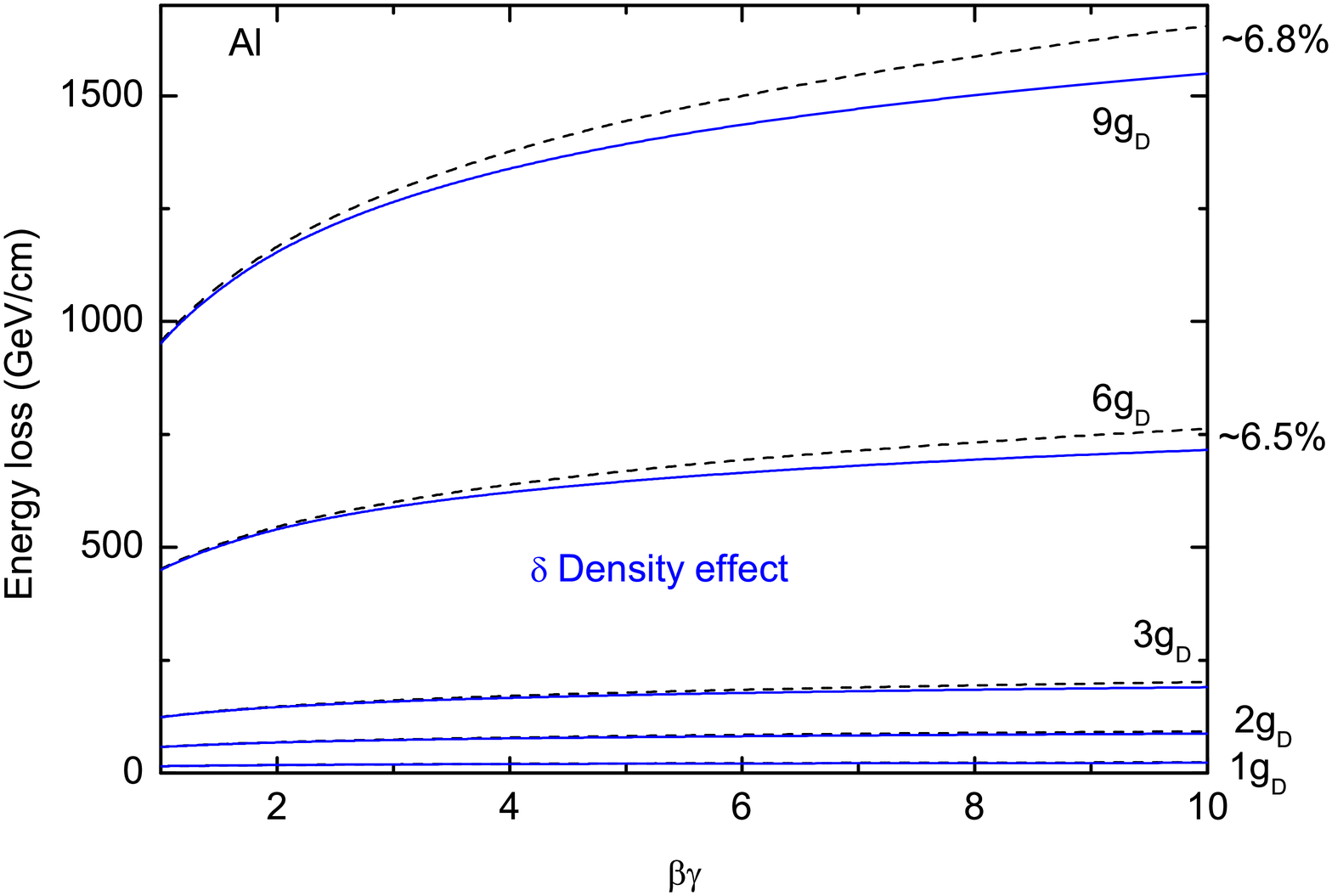}
%	\hspace{-0.2cm}
		\includegraphics[width=0.82\textwidth]{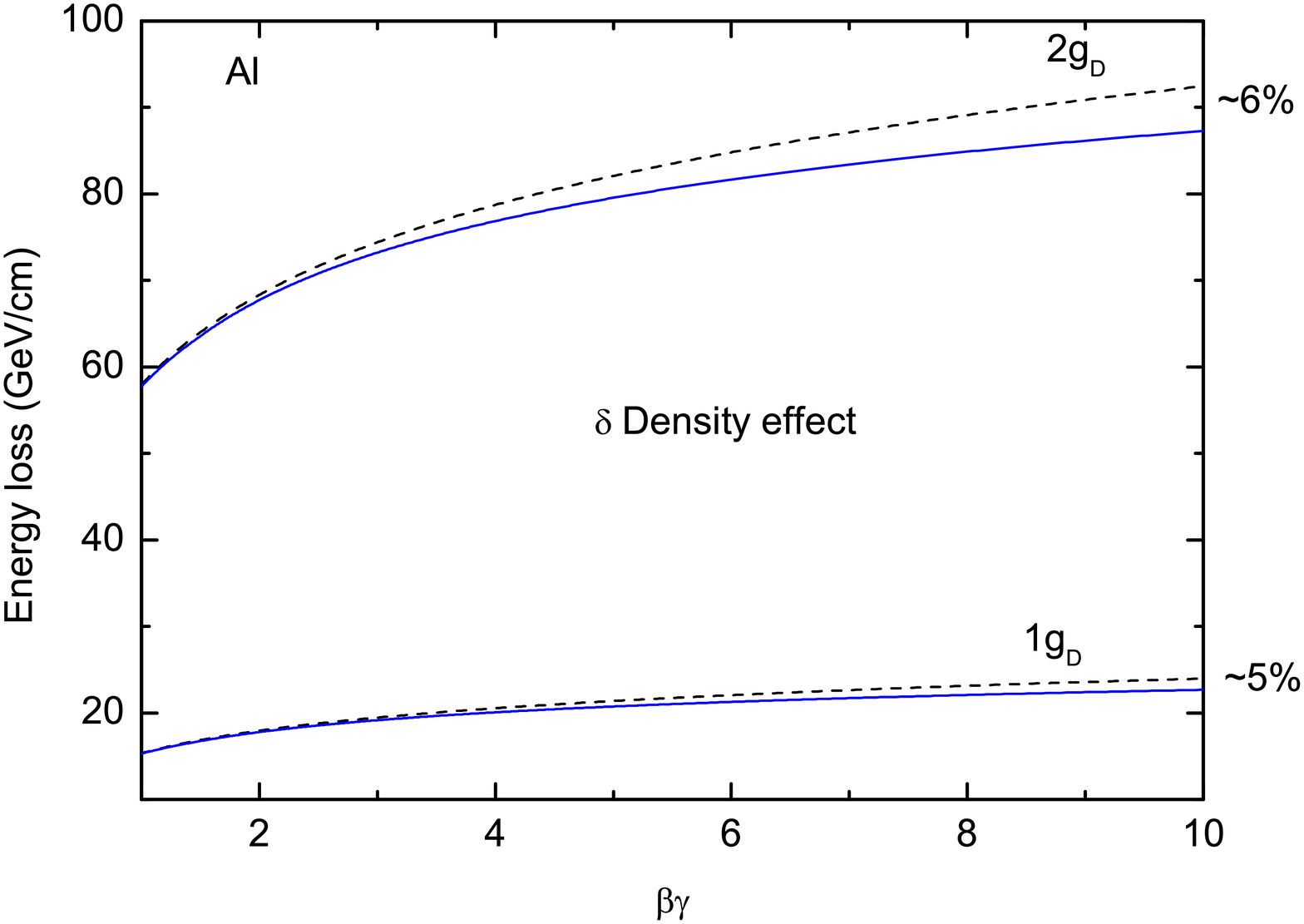}
	\end{center}
\vspace{-0.4cm}
	\caption{Energy loss of MMs in Aluminum vs $\beta\gamma$ computed with (blue curves) and without (black dashed curves) the density effect correction taken into account for $g= g_D$ and $2 g_D$  (bottom panel) and $g=3 g_D$ , $6g_D$ and $9g_D$ (top panel). Percentage values on the vertical scale (right) indicate the entity of the correction at $\beta\gamma \sim10$.}
	\label{fig:densityAlu2}
\end{figure*}

\begin{figure*}[htb!]
\vspace{1.0cm}
%	\hspace{-0.2cm}
	\begin{center}
		\includegraphics[width=0.82\textwidth]{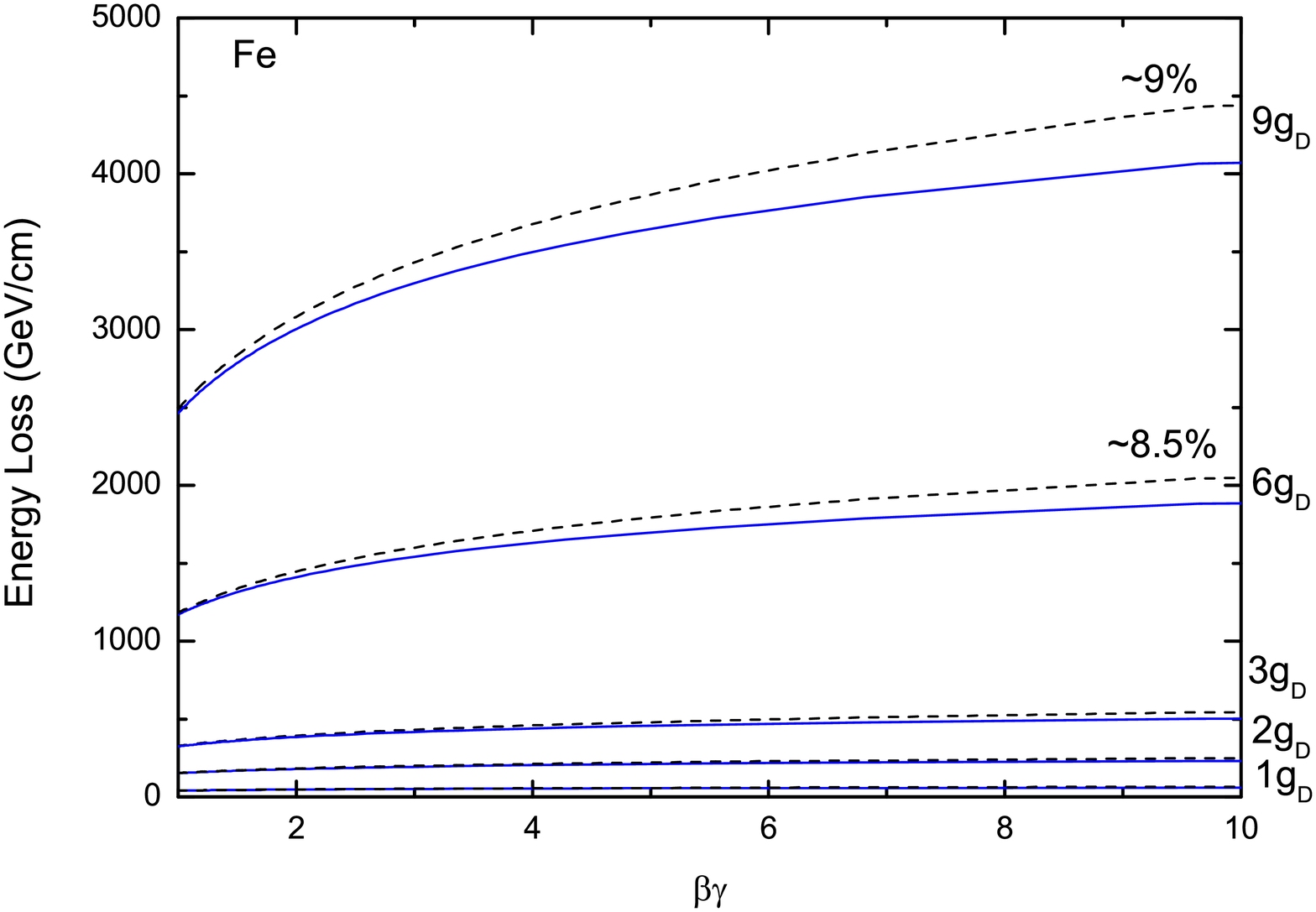}
%	\hspace{-0.2cm}
		\includegraphics[width=0.82\textwidth]{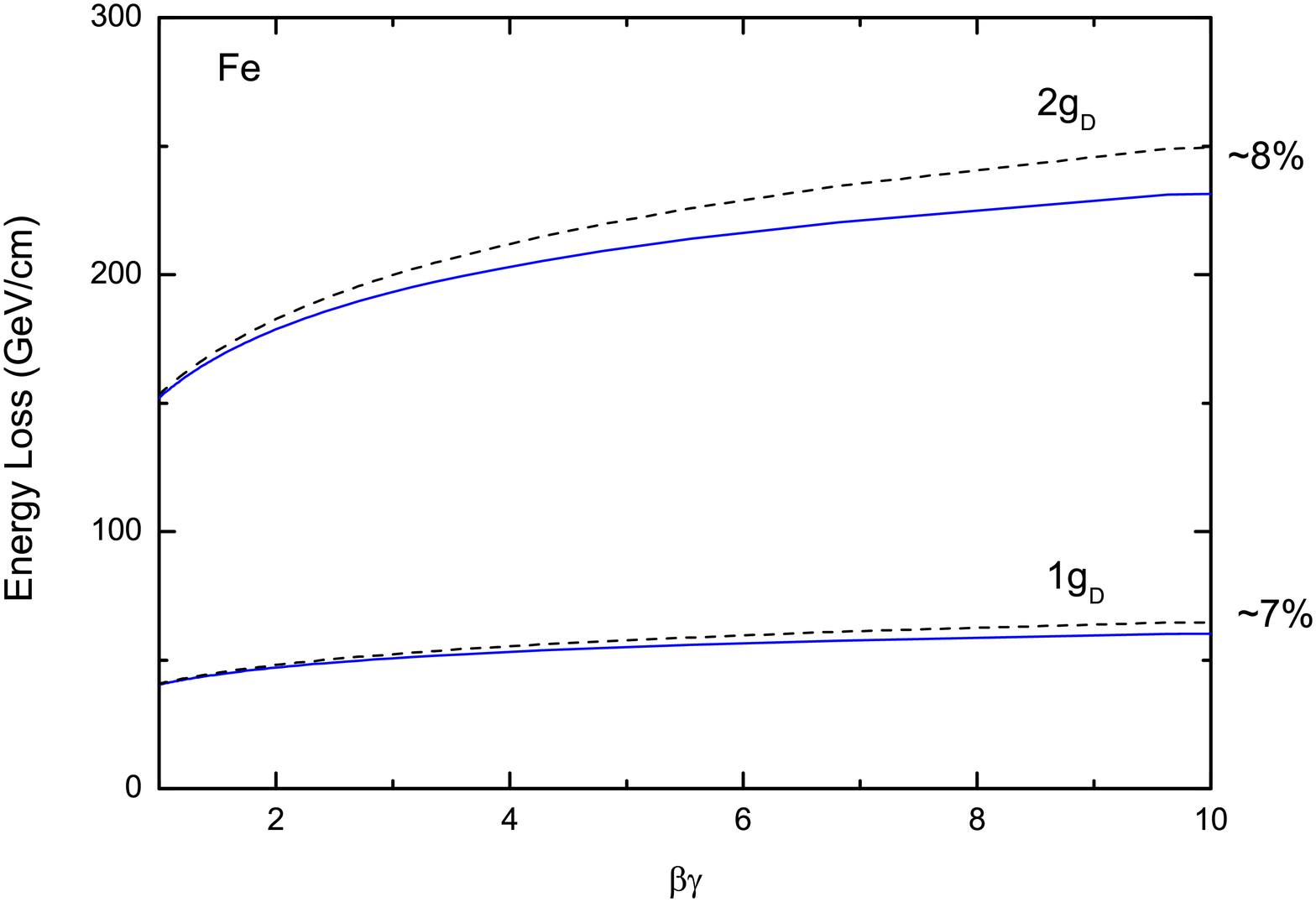}
	\end{center}
\vspace{-0.4cm}
	\caption{Energy loss of MMs in Iron vs $\beta\gamma$ computed with (blue curves) and without (black dashed curves) the density effect correction taken into account for $g= g_D$ and $2 g_D$  (bottom panel) and $g=3 g_D$ , $6g_D$ and $9g_D$ (top panel). Percentage values on the vertical scale (right) indicate the entity of the correction at $\beta\gamma \sim10$.}
	\label{fig:densityFe}
\end{figure*}

\begin{figure*}[tb!]
\vspace{1.0cm}
%	\hspace{-0.2cm}
	\begin{center}
		\includegraphics[width=0.82\textwidth]{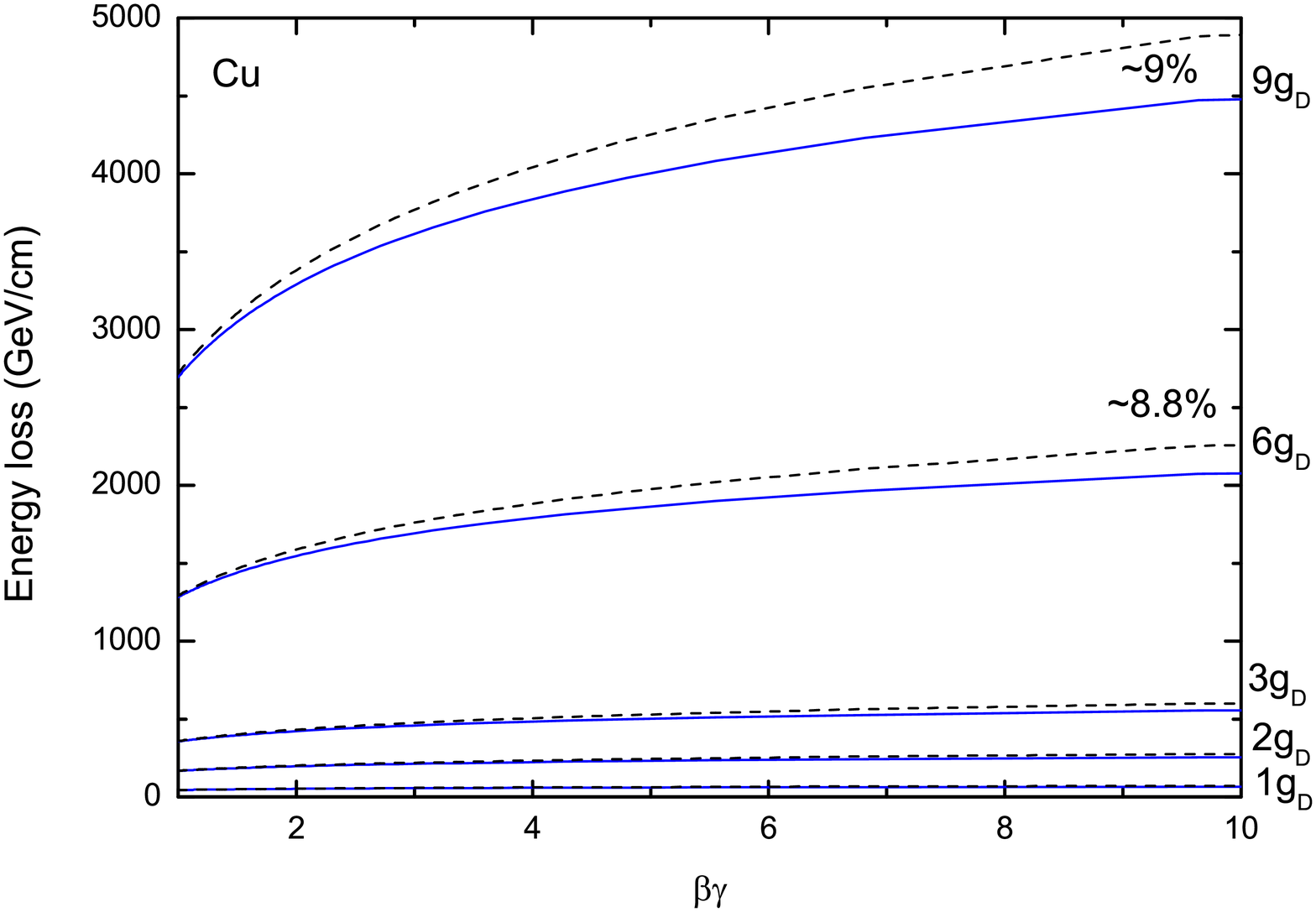}
%	\hspace{-0.2cm}
		\includegraphics[width=0.82\textwidth]{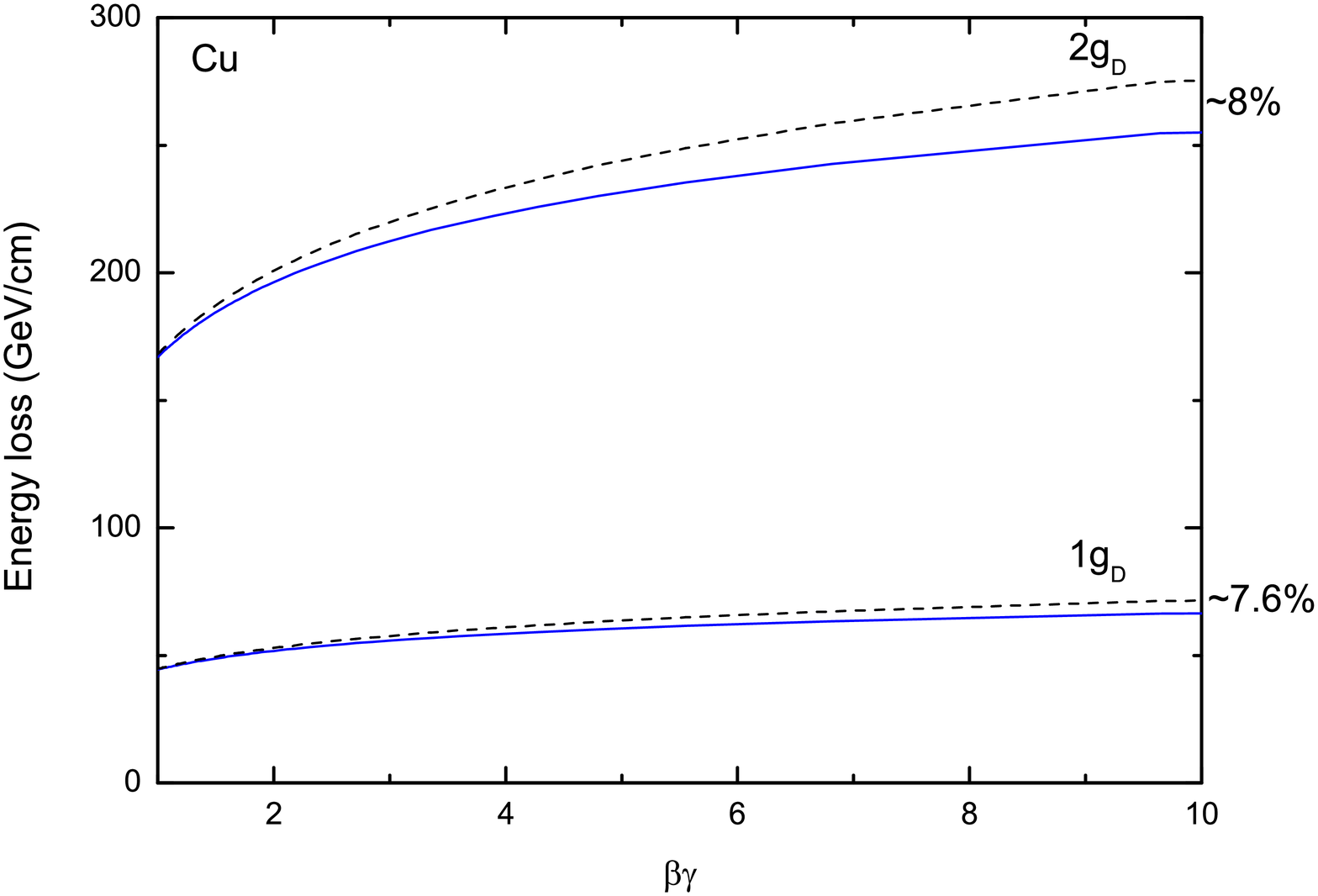}
	\end{center}
\vspace{-0.4cm}
	\caption{Energy loss of MMs in Copper vs $\beta\gamma$ computed with (blue curves) and without (black dashed curves) the density effect correction taken into account for $g= g_D$ and $2 g_D$  (bottom panel) and $g=3 g_D$ , $6g_D$ and $9g_D$ (top panel). Percentage values on the vertical scale (right) indicate the entity of the correction at $\beta\gamma \sim10$.}
	\label{fig:densityCu}
\end{figure*}

Figs.\ref{fig:densityAlu1}-\ref{fig:densityCu} show the contribution of the density effect on the MM energy losses for different values of the magnetic charge. At $\beta\gamma \sim10$ the correction in Aluminum (Iron) is $\sim5\% (7\%)$ for $g_D$ magnetic monopoles, and $\sim 7\% (9\%)$ for $g=9 g_D$ (Figs.\ref{fig:densityAlu2},\ref{fig:densityFe}).

\newpage

\newpage

\section{Energy losses of monopoles with $10^{-3} < \beta < 10^{-2}$}
\label{sec:medium}

For monopoles of medium velocity the energy losses can be computed assuming that the medium is a degenerate electron gas \cite{derkaoui1,Ahlen2}. Using the same notation as in Sec.\ref{sec:fast}, one has

\begin{equation}
\label{eq:medium}
%-\frac{dE}{dx} = \frac{2\pi \, N_{e} \,g^{2} \,e^{2}\beta}{m_{e}c \,v_F} \left[\ln\frac{2m_{e}v_F\Lambda}{\hbar} - 0.5\right]
-\frac{dE}{dx} = C \, \frac{Z}{A} \frac{c}{2v_F}\,g^{2}\, \beta  \left[\ln\frac{2m_{e}v_F\Lambda}{\hbar} - 0.5\right] \quad \textrm{MeVg$^{-1}$cm$^2$}
\end{equation}

\noindent where $v_F = (\hbar/m_e)(3\pi^2 N_e)^{1/3}$ is the Fermi velocity. $\Lambda = 53 \times 10^{-10}$ cm is the Bohr radius for non-conductors or bulk (non-valence) electrons in conductors.
In conductors, there are two contributions to the energy loss: from bulk electrons and from conduction electrons,

\begin{equation*}
-\frac{dE}{dx} = \left(-\frac{dE}{dx} \right)_{bulk} + \left(-\frac{dE}{dx} \right)_{conduction}
\end{equation*}

\noindent In this case $N_e$ is replaced by the density of bulk and conduction electrons, respectively. $\Lambda$ is the Bohr radius for the bulk contribution to the energy loss; $\Lambda=\frac{50\,  a\, T_m}{T}$ for the conduction electrons' contribution. $a=\sqrt[3]{A/(N_A\times\rho)}$ , $T_m$ is the target fusion temperature, and T is the temperature.\\

In Table \ref{tab:param} are given the densities of conduction and bulk electrons, Fermi velocities and fusion temperature for Aluminum, Iron and Copper.

\begin{table}[h!]
\centering
\begin{tabular}{|c|c|c|c|c|c|}
\hline
Material & $N_{eCond}$ (cm$^{-3}$) & $N_{eBulk}$  (cm$^{-3}$) & $v_{FCond}$ (cm/s) & $v_{FBulk}$ (cm/s)   & $T_m$ (K)\\
\hline
Al  & $18.2 \times 10^{22}$ &   $6.011 \times 10^{23}$   &   $2.03 \times 10^8$	  &   $3.024 \times 10^8$  &   933.52 \\
\hline
Fe & $17 \times 10^{22}$     &  $2.03 \times 10^{24}$     &   $1.98 \times 10^8$	  &   $4.54 \times 10^8$    &   1811 \\
\hline
Cu & $8.47 \times 10^{22}$ &  $2.38 \times 10^{24}$     &   $1.57 \times 10^8$	  &   $4.78 \times 10^8$    &   1358  \\
\hline
%Si  & 0 & $6.99 \times 10^{23}$ & 0	& $3.18\times 10^8$  cm/s$   &   933.5\\
%\hline
\end{tabular} 
\caption{Conduction and bulk electron density ($N_{eCond}$, $N_{eBulk}$, respectively), Fermi velocity ($v_{FCond}$ and $v_{FBulk}$, respectively) and fusion temperature for Aluminum, Iron and Copper.}
\label{tab:param}
\end{table}

\noindent For Aluminum
\begin{equation*}
-\frac{dE}{dx}(Al)= (13.7+80) \times n^2 \beta \quad(\textrm{GeVg$^{-1}$cm$^2$})
\end{equation*}

\noindent This result is not too different from the estimate in G. Giacomelli \cite{prof}
\begin{equation*}
-\frac{dE}{dx}(Al)= (20+130) \times n^2 \beta \quad(\textrm{GeVg$^{-1}$cm$^2$})
\end{equation*}

\noindent For Iron,
\begin{equation*}
-\frac{dE}{dx}(Fe)= (18.9+28.4) \times n^2 \beta \quad(\textrm{GeVg$^{-1}$cm$^2$})
\end{equation*}

\noindent which is consistent with the expression derived from J. Derkaoui et al. \cite{derkaoui1}  when the same values of density and temperature are considered.\\

\noindent For Copper

\begin{equation*}
-\frac{dE}{dx}(Cu)= (19.45 + 14.54) \times n^2 \beta \quad(\textrm{GeVg$^{-1}$cm$^2$})
\end{equation*}

%\noindent Silicon is a non conductor so the energy loss is due only to the bulk electrons; $N_{eBulk}= 6.99 \times 10^{23}  \textrm{cm}^{-3}$ ; $v_{FBulk} = 3.18 \times 10^8$  cm/s.
%\begin{equation*}
%-\frac{dE}{dx}(Si)=(19.26) \times n^2 \beta \quad (\textrm{GeVg$^{-1}$cm$^2$})
%\end{equation*}

\newpage

\section{Energy losses of slow monopoles at $\beta< 10^{-5}$}
\label{sec:slow}
In paramagnetic materials the energy loss is given by \cite{bracci}

\begin{equation}
-\frac{dE}{dx}= \mu\times\frac{4\pi\hbar \,g \, e\, N}{c\, m_e}\times 0.6
\end{equation}

\noindent $\mu$ is the magnetic moment of the target atoms in Bohr magneton and $N$ is their density.\\
\noindent For Aluminum $\mu$ = 3.6415; \\
\noindent for Iron $\mu$ = 2.201;  \\
\noindent for Copper, $\mu$ = 2.2233.\\

\begin{figure}[htb!]
          \vspace{0.5cm}
	\begin{center}
		\includegraphics[width=0.8\textwidth]{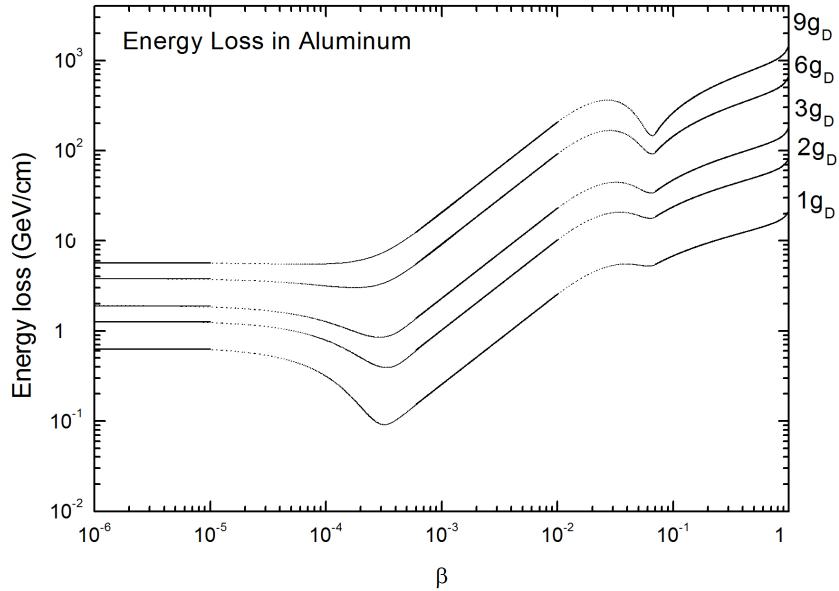}
	\end{center}
           \vspace{-0.4cm}
	\caption{Energy loss of magnetic monopoles with charge $1g_D, 2g_D, 3g_D, 6g_D$ and $9g_D$ in Aluminum.}
	\label{fig:elossAl}
\end{figure}

\begin{figure}[htb!]
%	\vspace{0.5cm}
	\begin{center}
		\includegraphics[width=0.8\textwidth]{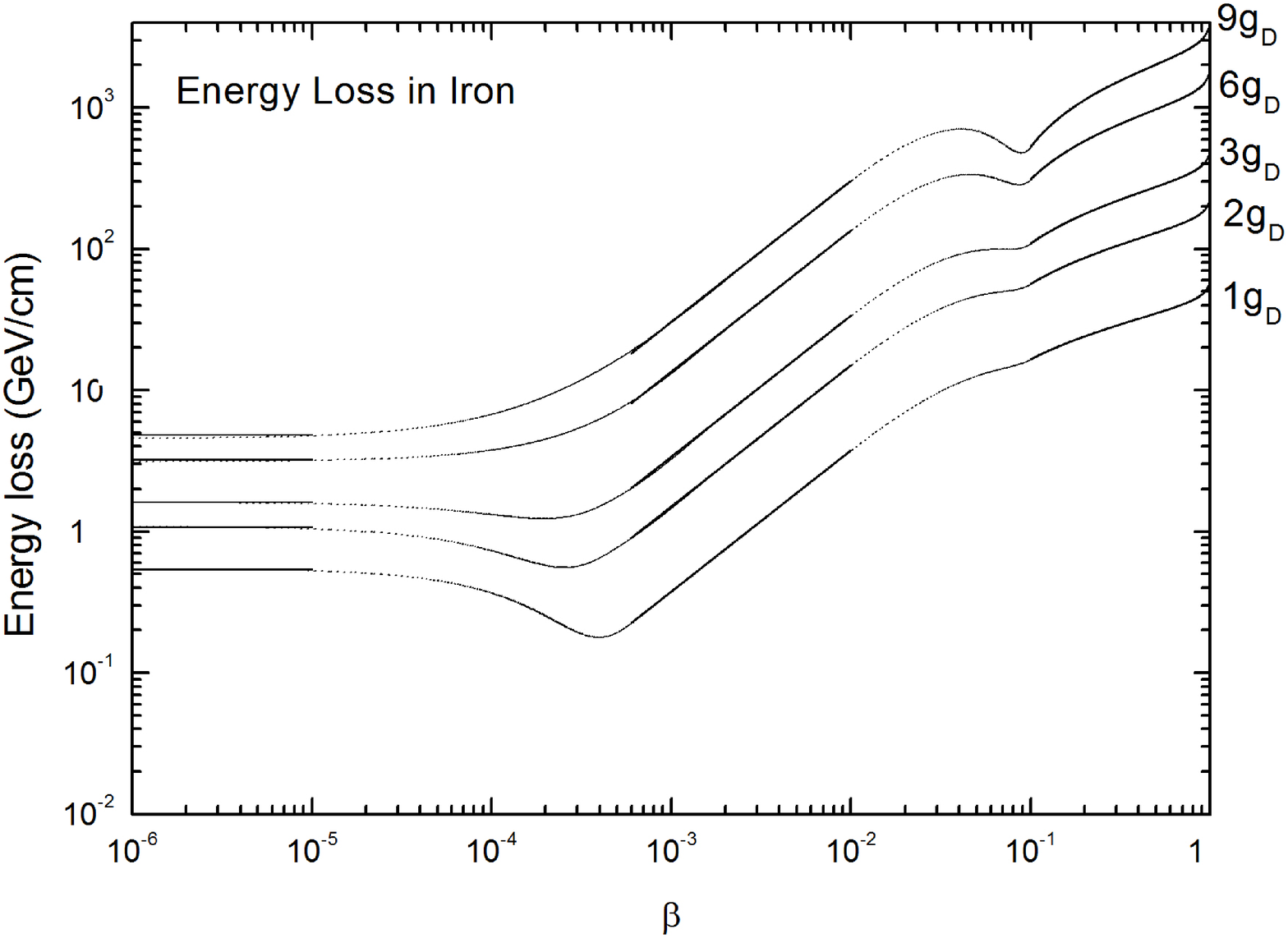}
	\end{center}
	\vspace{-0.4cm}
	\caption{Energy loss of magnetic monopoles with charge $1g_D, 2g_D, 3g_D, 6g_D$ and $9g_D$ in Iron.}
	\label{fig:elossFe}
\end{figure}

\begin{figure}[h!]
	\vspace{0.5cm}
	\begin{center}
		\includegraphics[width=0.8\textwidth]{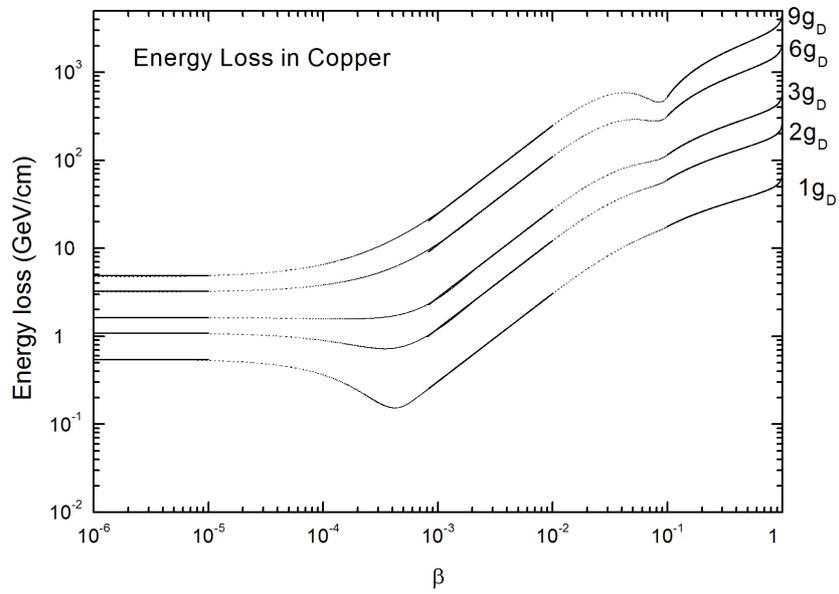}
	\end{center}
	\vspace{-0.4cm}
	\caption{Energy loss of magnetic monopoles with charge $1g_D, 2g_D, 3g_D, 6g_D$ and $9g_D$ in Copper.}
	\label{fig:elossCu}
\end{figure}

Figs. \ref{fig:elossAl}, \ref{fig:elossFe} and \ref{fig:elossCu} show the energy losses versus $\beta$ for monopoles with magnetic charges $1g_D, 2g_D, 3g_D, 6g_D$ and $9g_D$ in Aluminum, Iron and Copper (dotted lines are rough interpolations).

\section{The Range of Magnetic Monopoles}
\label{sec:range}
The range R of magnetic monopoles was obtained by integrating the energy loss :
\begin{equation}
\label{eq:range}
R=\int_{E_{min}}^{E_0} \frac{dE}{dE/dx}
\end{equation}

\noindent In the formula $E_0$  is the pole initial kinetic energy and $E_{min}$ is the kinetic energy corresponding to $\beta\sim10^{-6}$.\\ 

\begin{figure}[htb!]
	\vspace{-0.3cm}
	\hspace{-2cm}
%	\begin{center}
		\includegraphics[width=0.63\textwidth]{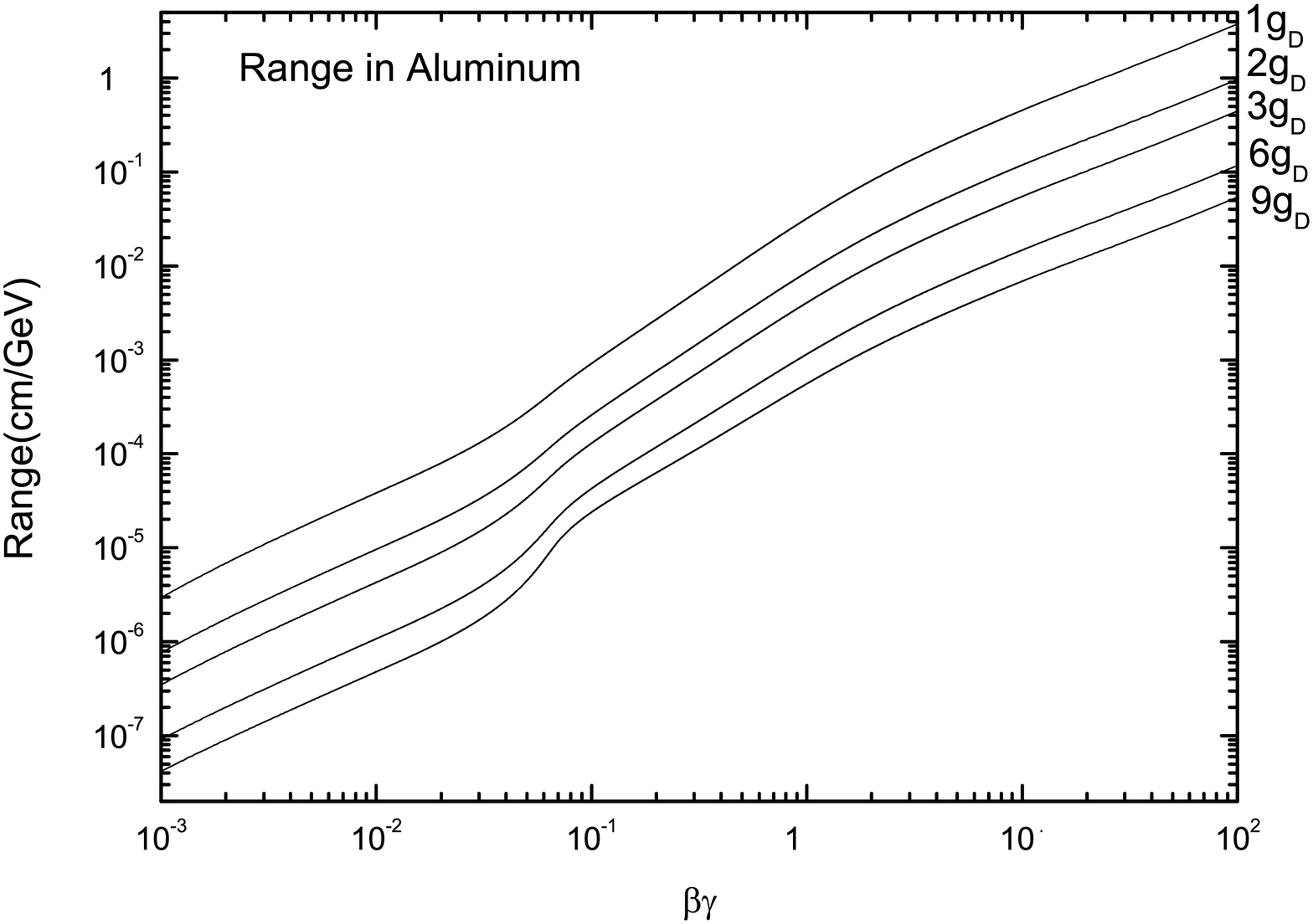}
                     \includegraphics[width=0.63\textwidth]{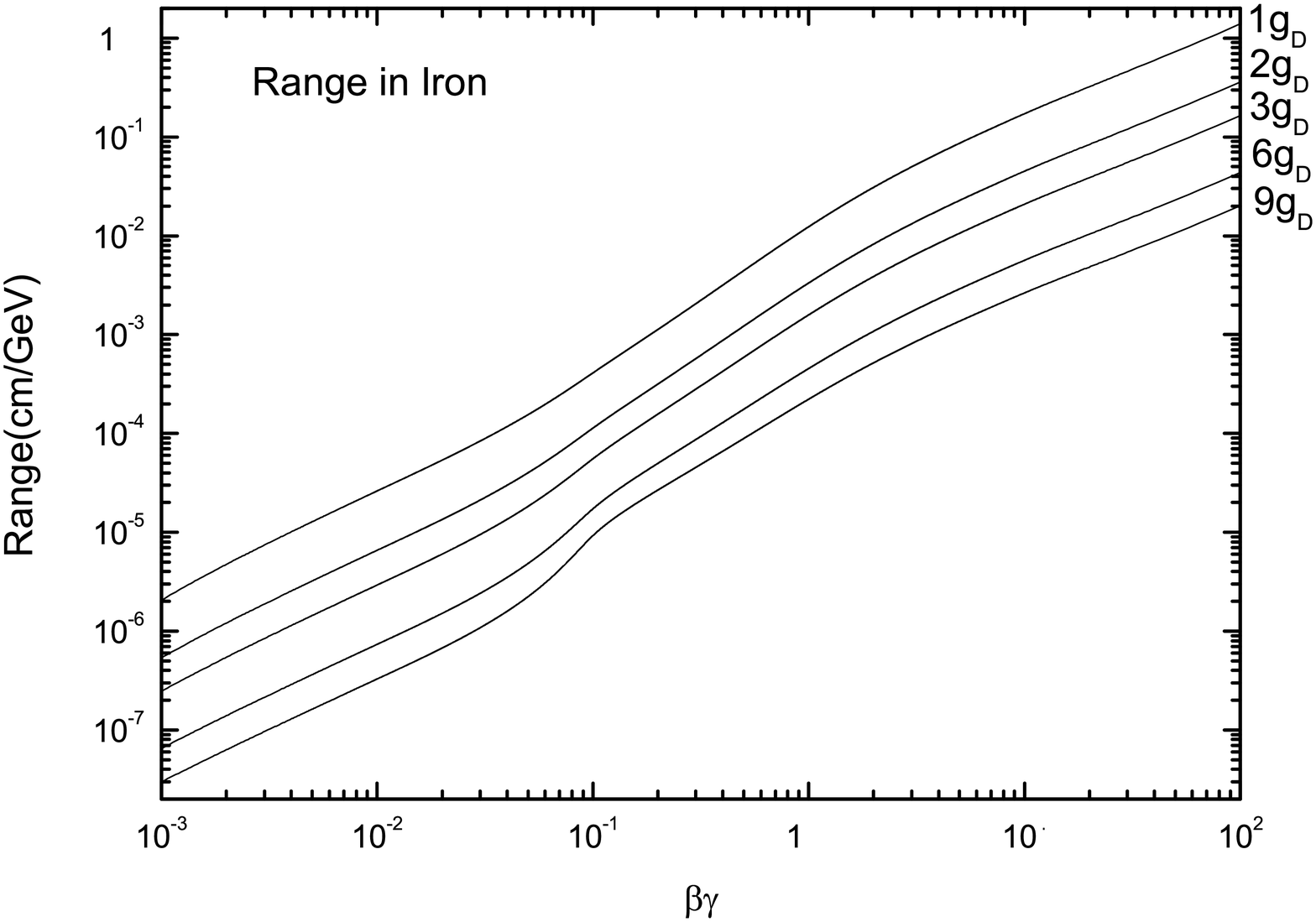}
	\vspace{0.5cm}
	\hspace{-2cm}
%	\begin{center}
                     \includegraphics[width=0.63\textwidth]{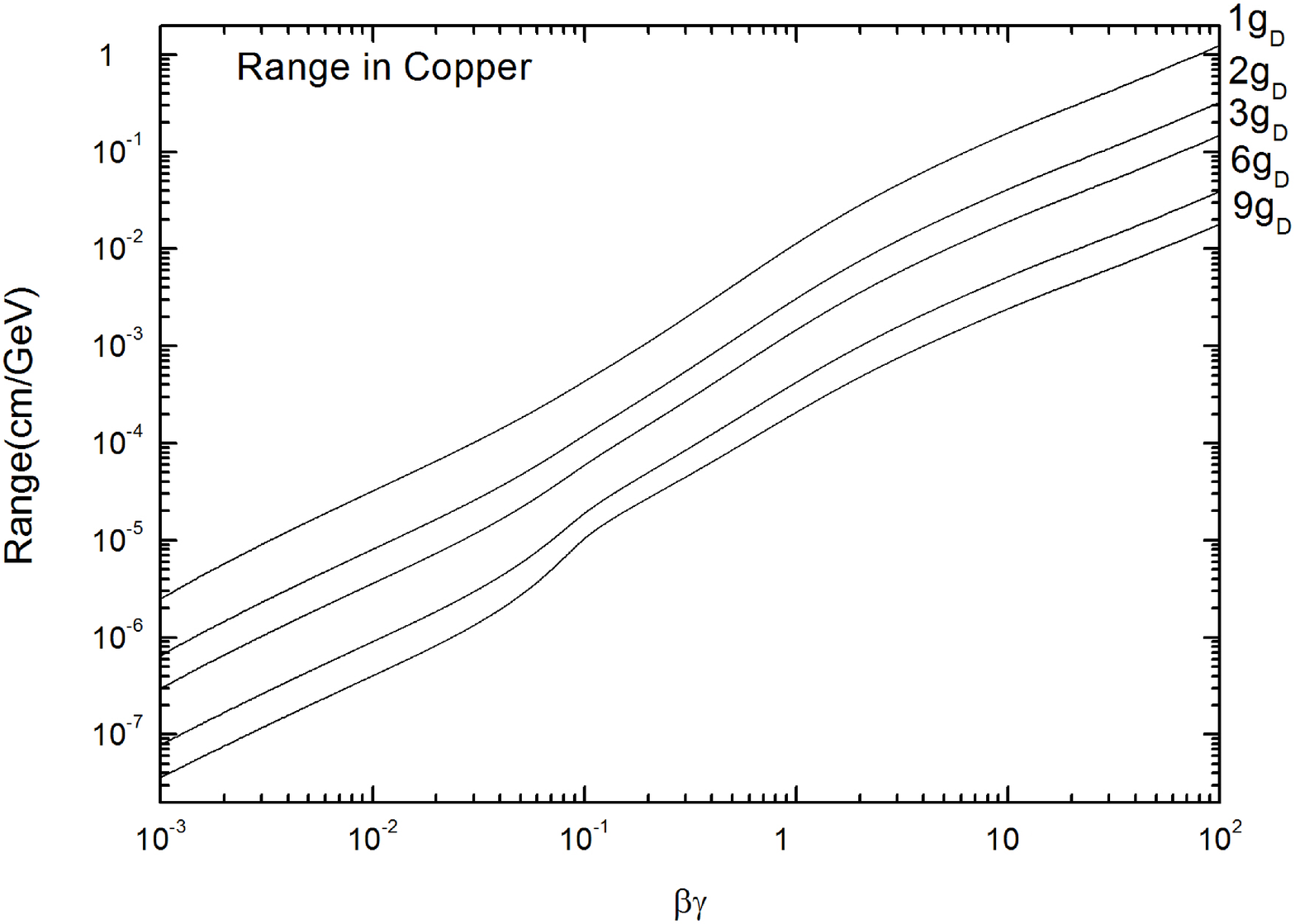}
%	\end{center}
	\vspace{-0.4cm}
	\caption{Range/Mass of $1g_D, 2g_D, 3g_D, 6g_D$ and $9g_D$ MMs in Aluminum (top left), Iron (top right) and Copper (bottom).}
	\label{fig:rangeAl0}
\end{figure}

In Fig.\ref{fig:rangeAl0} the curves show the Range/Mass (cm/GeV) versus $\beta\gamma$ of MMs with magnetic charge $1g_D, 2g_D, 3g_D, 6g_D$ and $9g_D$ in Aluminum, Iron and Copper, respectively.\\

\begin{figure}[h!]
%	\begin{center}
	\vspace{1cm}
	\hspace{-2.cm}
		\includegraphics[width=0.65\textwidth]{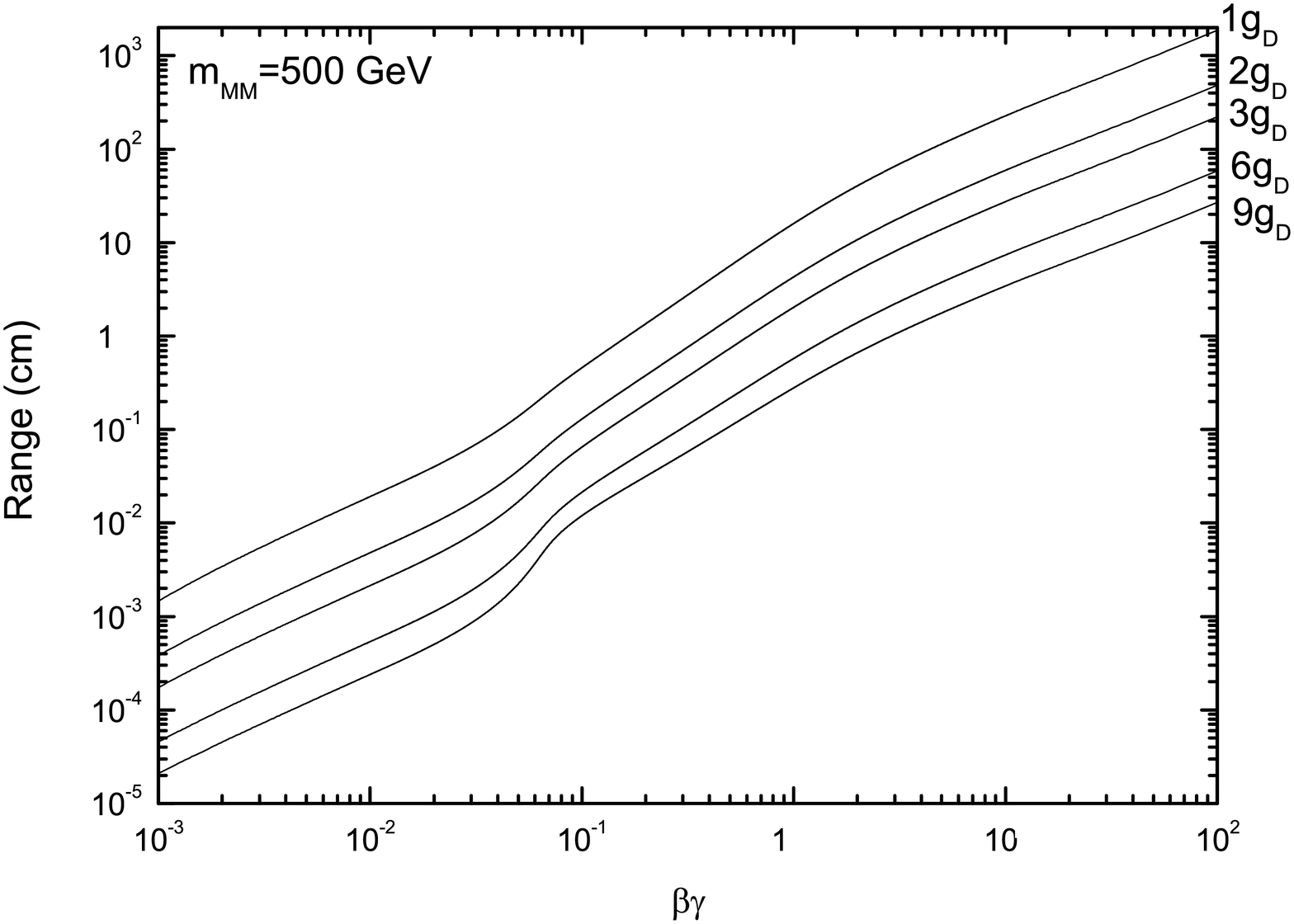}
%           \hspace{0.5cm}
		\includegraphics[width=0.65\textwidth]{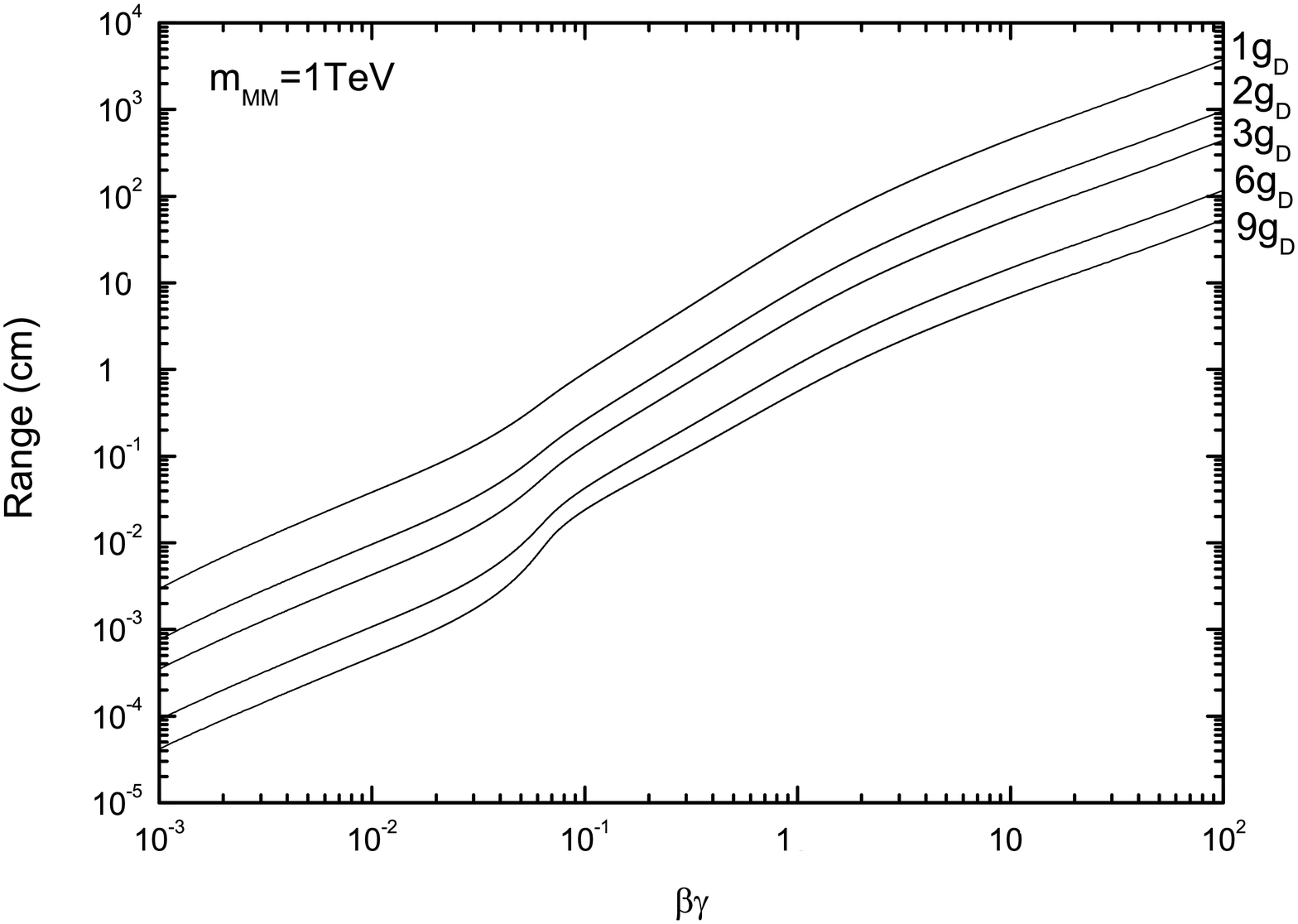}
%	\begin{center}
	\vspace{0.5cm}
	\hspace{-2.cm}
		\includegraphics[width=0.65\textwidth]{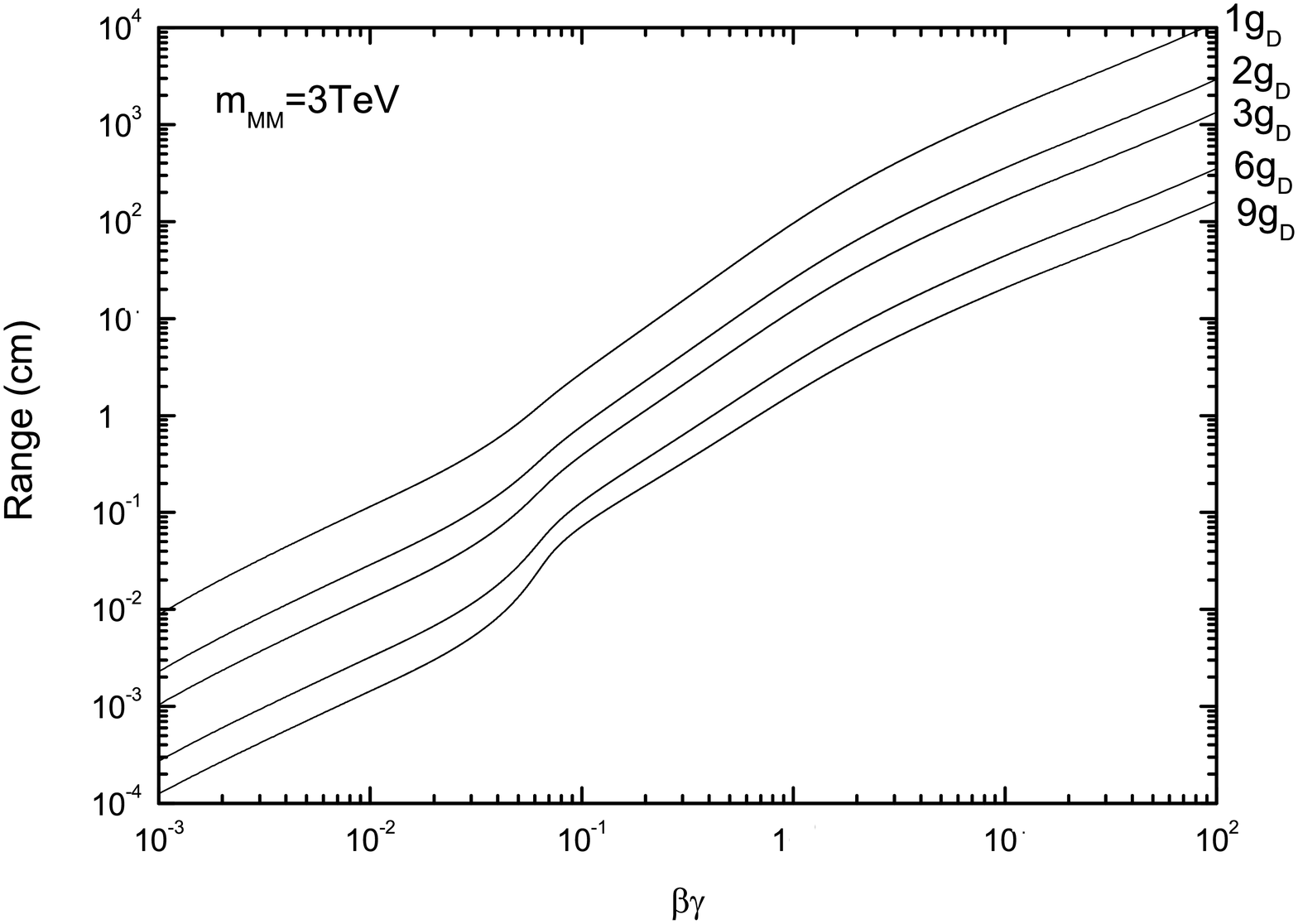}
%	\end{center}
	\caption{Ranges versus $\beta\gamma$ in Aluminum for MMs with mass 0.5, 1 and 3 TeV and different magnetic charges ($1g_D, 2g_D, 3g_D, 6g_D$ and $9g_D$).}
	\label{fig:rangeAl1}
\end{figure}

\begin{figure}[[htb!]
%	\begin{center}
	\vspace{5cm}
	\hspace{-2.cm}
		\includegraphics[width=0.65\textwidth]{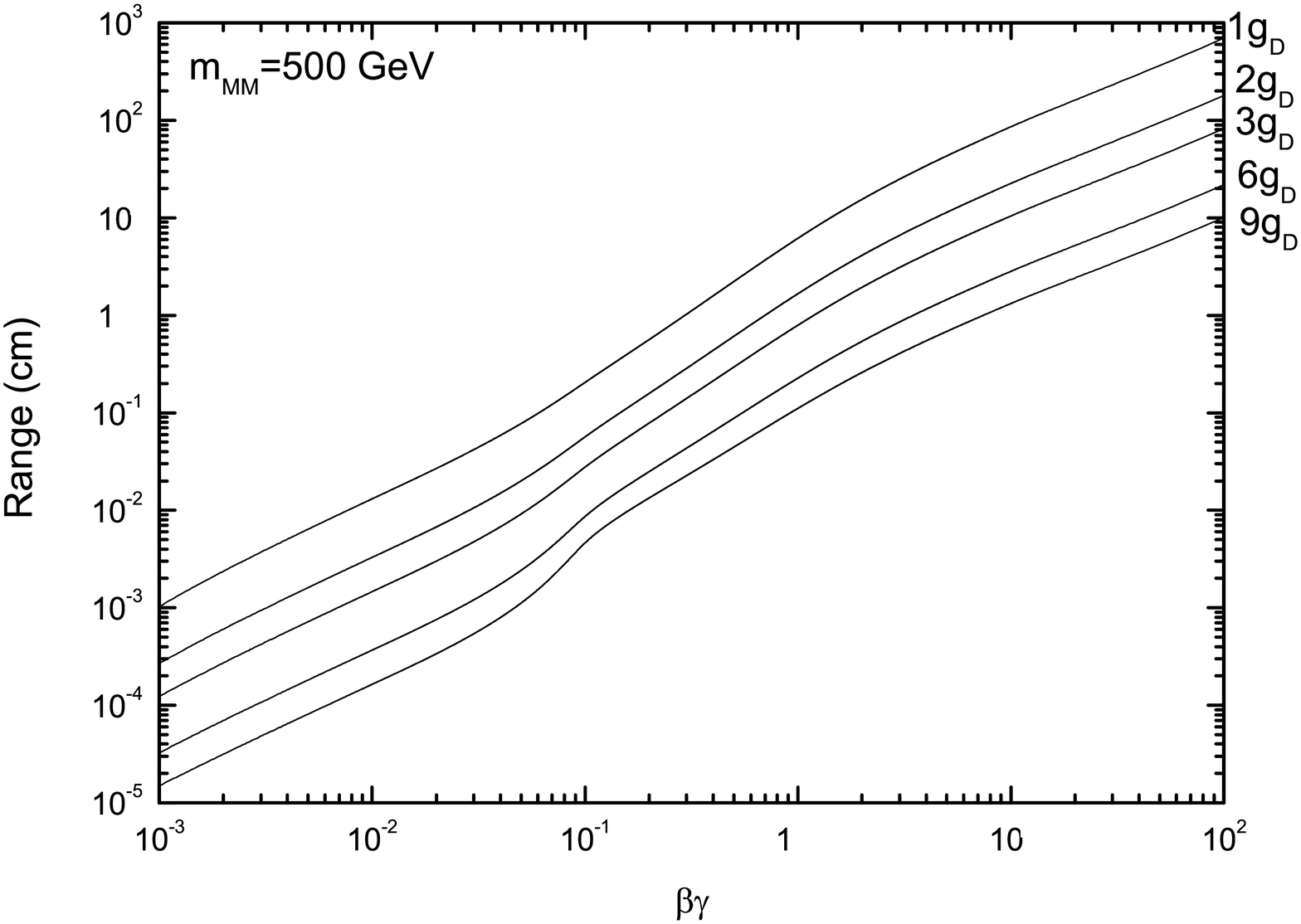}
		\includegraphics[width=0.65\textwidth]{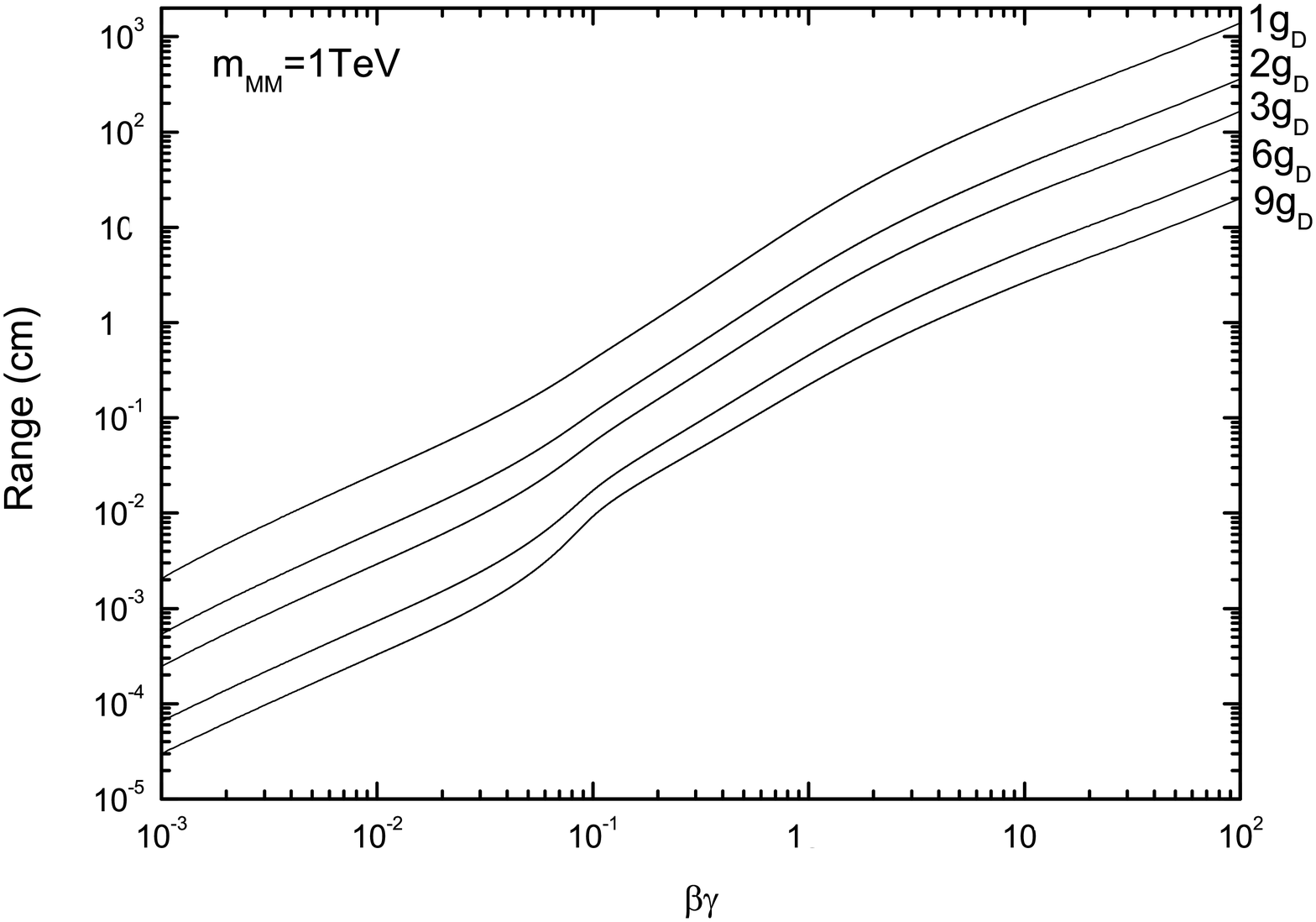}
%	\begin{center}
	\vspace{0.5cm}
	\hspace{-2.cm}
 		\includegraphics[width=0.65\textwidth]{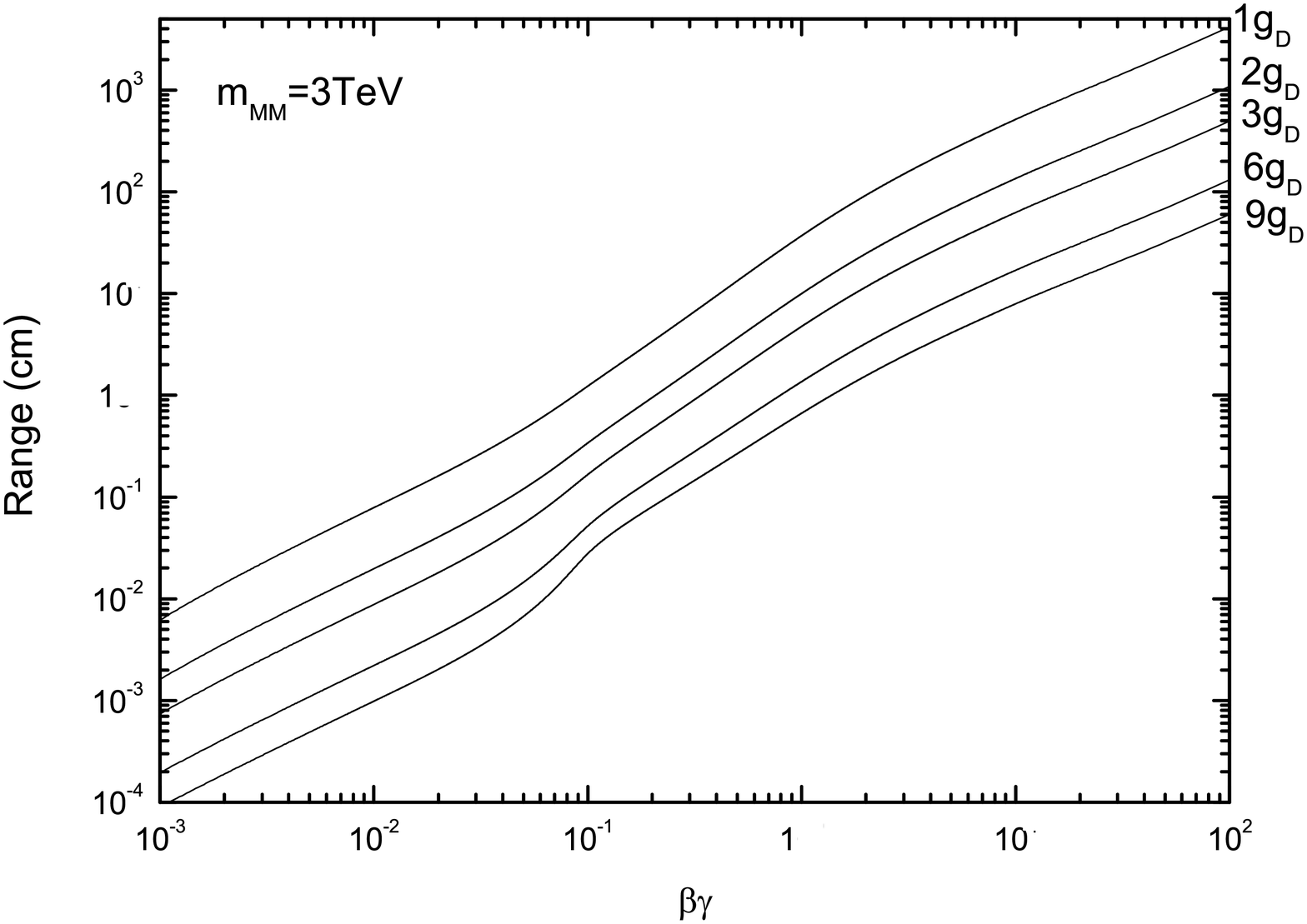}
%	\end{center}
	\vspace{-0.4cm}
	\caption{Ranges versus $\beta\gamma$ in Iron for MMs with mass 0.5, 1 and 3 TeV and different magnetic charges ($1g_D, 2g_D, 3g_D, 6g_D$ and $9g_D$).}
	\label{fig:rangeFe1}
\end{figure}

\begin{figure}[htb!]
%	\begin{center}
	\vspace{5cm}
	\hspace{-2.cm}
		\includegraphics[width=0.65\textwidth]{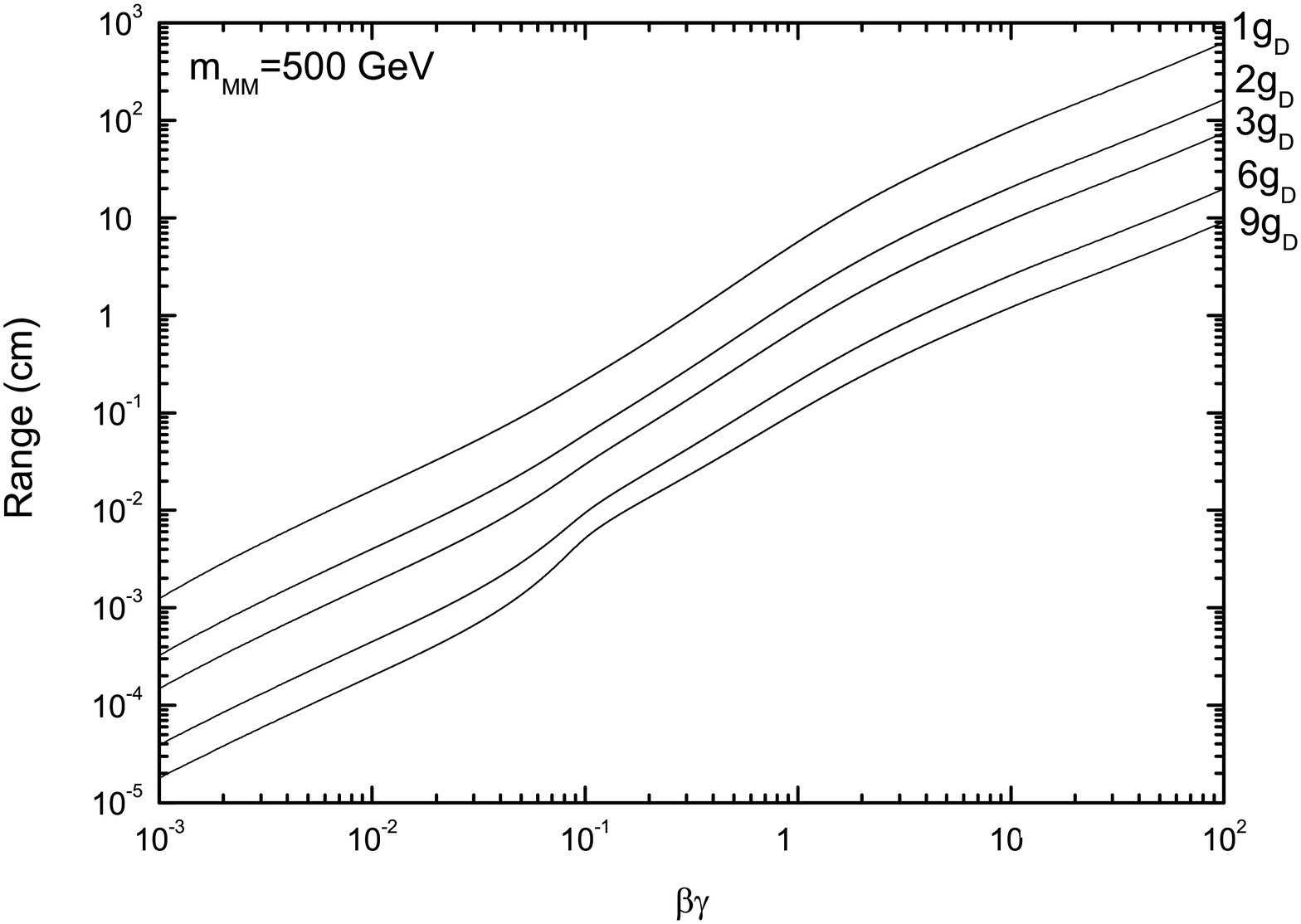}
		\includegraphics[width=0.65\textwidth]{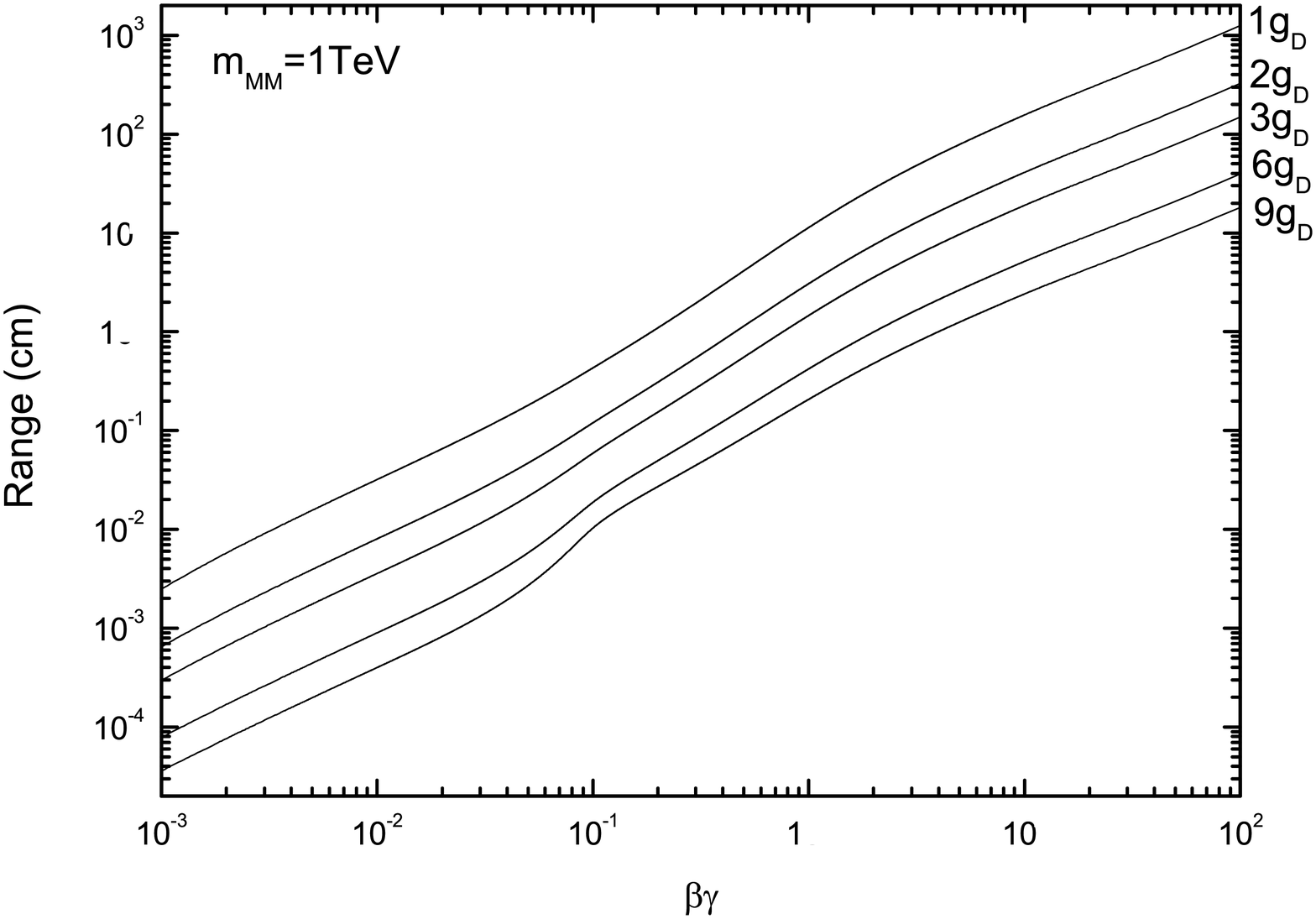}
	\vspace{0.5cm}
	\hspace{-2.cm}
%	\begin{center}
		\includegraphics[width=0.65\textwidth]{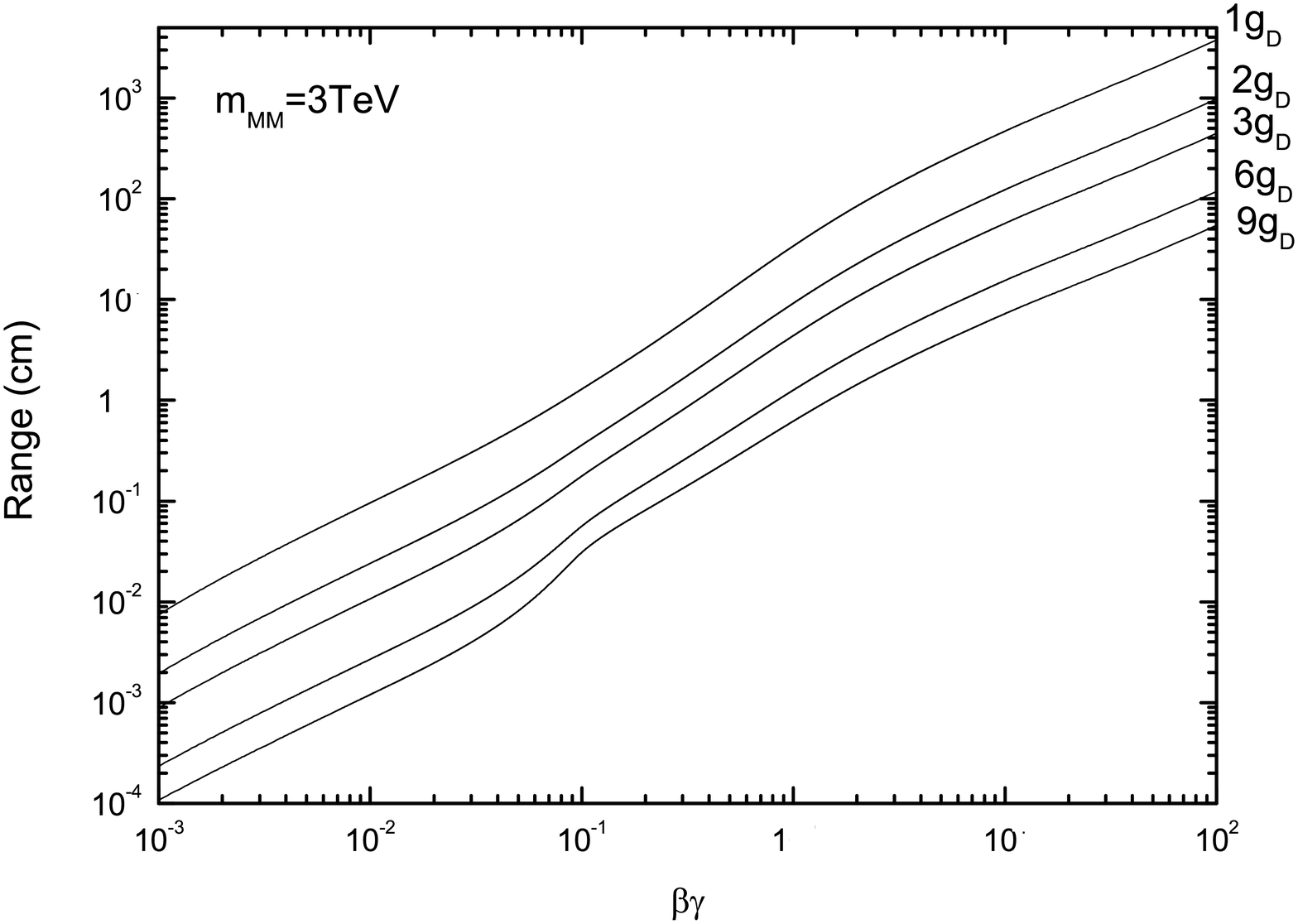}
%	\end{center}
	\vspace{-0.4cm}
	\caption{Ranges  in Copper versus $\beta\gamma$ for MMs with mass 0.5, 1 and 3 TeV and different magnetic charges ($1g_D, 2g_D, 3g_D, 6g_D$ and $9g_D$).}
	\label{fig:rangeCu1}
\end{figure}

The ranges versus $\beta\gamma$ of MMs with magnetic charge  $1g_D, 2g_D, 3g_D, 6g_D$ and $9g_D$ and mass 0.5, 1 and 3 TeV in Aluminum, Iron and Copper are plotted in Figs.\ref{fig:rangeAl1}, \ref{fig:rangeFe1} and \ref{fig:rangeCu1}, respectively. 

\newpage 

\newpage

\newpage

\section{Magnetic Monopoles in Nuclear Track Detectors (NTDs)}
\label{sec:ntd}

\begin{figure}[htb!]
	\vspace{1cm}
	\begin{center}
		\includegraphics[width=0.8\textwidth]{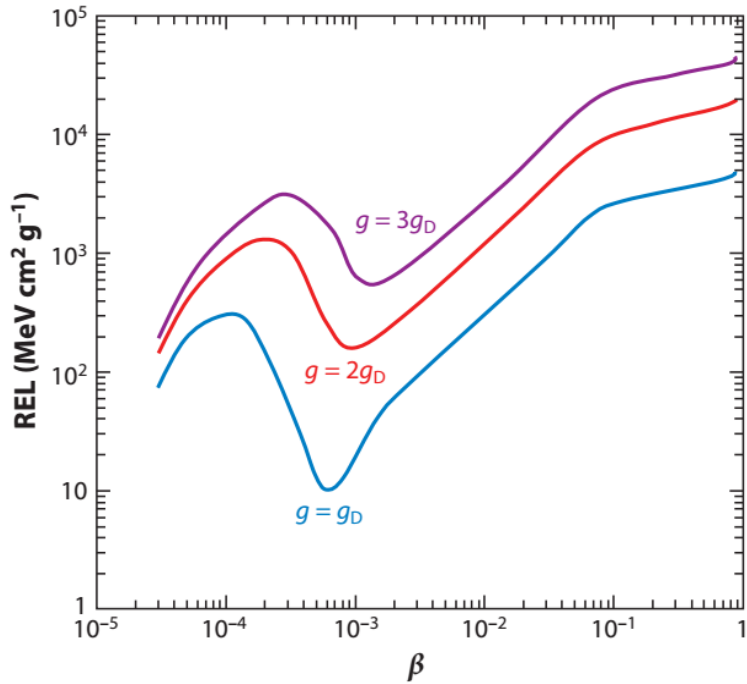}
	\end{center}
	\vspace{-0.4cm}
	\caption{Restricted energy loss (REL) in CR39\small{\textregistered} vs $\beta$ for $1g_D, 2g_D$ and $3g_D$  MMs. \cite{AnRev}}
	\label{fig:rel}
\end{figure}

Nuclear-track detectors (NTDs) are sensitive to the energy deposited in a narrow cylindrical region (O(10 nm)) along the particle trajectory. \\
In polymers such as CR39\small{\textregistered} and Makrofol/Lexan the restricted energy loss (REL) is the relevant quantity.  At  $\beta > 0.05$, only  energy transfers yielding $\delta$-rays of energy $<200$ eV (350 eV) in CR39 (Makrofol/Lexan) contribute to REL. At lower velocities ($\beta < 10^{-2}$), elastic recoils from diamagnetic interactions between the monopole and atoms have to be considered. Atomic elastic recoils give rise to a bump in the REL below $\beta\sim 10^{-3}$.
The Restricted Energy Loss (REL) of monopoles with charges $g=g_D, 2g_D$ and $3 g_D$ in CR39\small{\textregistered} is shown in Fig.\ref{fig:rel} (from \cite{AnRev}; detailed computation in \cite{derkaoui2}). 

\end{document}